\documentclass[runningheads]{llncs}
\usepackage{graphicx}

\usepackage{amsmath}
\usepackage{graphicx}%
\usepackage{amsfonts}%
\usepackage{amssymb}

\newcommand{\eqnb}{\begin{equation}}
\newcommand{\eqne}{\end{equation}}

\newtheorem{The}{Theorem}

\begin{document}
\title{A Stochastic Model for File Lifetime and Security in Data Center Networks\thanks{Quan-Lin Li was supported by the National Natural Science Foundation of China under grant No. 71671158 and No. 71471160, and by the Natural Science Foundation of Hebei province under grant No. G2017203277.}}
%
%
\author{Quan-Lin Li \and
Fan-Qi Ma \and Jing-Yu Ma}
\authorrunning{Q. L. Li et al.}
%
\institute{School of Economics and Management Sciences, Yanshan University, Qinhuangdao 066004, China\\
\email{liquanlin@tsinghua.edu.cn}}
\maketitle              
\begin{abstract}
Data center networks are an important infrastructure in various applications of modern information technologies.
Note that each data center always has a finite lifetime, thus once a data center
fails, then it will lose all its storage files and useful information.
For this, it is necessary to replicate and copy each important file into other data centers such that
this file can increase its lifetime of staying in a data center network.
In this paper, we describe a large-scale data center network with a file $d$-threshold policy,
which is to replicate each important file into at most $d-1$ other data centers
such that this file can maintain in the data center network under a given level of data security in the long-term.
To this end, we develop three relevant Markov processes to propose two effective methods for
assessing the file lifetime and data security.
By using the $RG$-factorizations, we show that the two methods are used to
be able to more effectively evaluate the file lifetime of large-scale data center networks.
We hope the methodology and results given in this paper are applicable in the file lifetime study of more
general data center networks with replication mechanism.

\keywords{Data center \and replication mechanism \and file lifetime \and data security \and Markov process
\and $RG$-factorization.}
\end{abstract}
\section{Introduction}

Data center networks are an important infrastructure for various applications of modern information technologies.
However, not only does each data center have a finite lifetime, but it is also possible to suffer from natural disasters
and man-made damages. Once a data center fails, then it will lose all its storage files and useful information.
In practice, such a data center failure has caused key innovation in design and management of large-scale data centers,
one of which is how to keep that no important file is lost in the data center network in the long-term. To do this, it is necessary to increase file and data availability
as far as possible. In this case, a file copy redundancy technology is developed as a simple mode that each important file is replicate into
a subset of data centers according to a comprehensive consideration for departments and/or geographical environment,
together with emergency responses for natural disasters and man-made damages. Therefore, during the last decade considerable attention
has been paid to developing stochastic model analysis for assessing file lifetime and data security in large-scale data center networks
with file replication mechanism. Also see Picconi et al. \cite{Pic:2007a} and \cite{Pic:2007b}, Kersch and Szabo \cite{Ker:2010}
and Feuillet and Robert \cite{Feu:2014} for more details.

Little work has been done on how to establish stochastic models (e.g., Markov processes, queueing theory and stochastic game) to assess the file
lifetime and data security in a large-scale data center network with file replication mechanism. Intuitively, such a study is more interesting,
difficult and challenging due to the fact that the mathematical modelling is based on reliability and security analysis of large-scale
stochastic networks. Based on this, important topics include
failure prediction for data centers, how to control new data centers joining this network, how to design and optimize file
replication mechanism. In addition, there are still some interesting issues, such as cost analysis of recovering lost data, effect of file replication mechanism
and bandwidth limitation, durability and availability of data, and how to control the file lost probability.
Readers may refer to recent publications for details, among which,
{\it data storage systems} by Blake and Rodrigues \cite{Bla:2003}, Utard and Vernois \cite{Uta:2004},
Lian et al. \cite{Lia:2005}, Chun \cite{Chu:2006} and
Ramabhadran and Pasquale \cite{Ram:2006}, \cite{Ram:2008} and \cite{Ram:2010};
{\it DHT replication} by Picconi et al. \cite{Pic:2007a} and \cite{Pic:2007b}, Kersch and Szabo \cite{Ker:2010},
Pace et al. \cite{Pac:2011} and Kniesburges et al. \cite{Kni:2013};
{\it failure prediction} by Pinheiro et al. \cite{Pin:2007}; and
{\it large-scale stochastic networks with unreliable processors} by Feuillet and Robert \cite{Feu:2014},  Sun et al. \cite{Sun:2016}
and Aghajani et al. \cite{Agh:2018}.

The main contributions of this paper are twofold. The first one is
to describe a large-scale data center network, in which each data center
may fail and new data centers can join this network, and a file $d$-replication policy is
proposed to increase file lifetime and data security.
The second one is to develop three relevant Markov processes,
which lead to two effective methods for assessing the file lifetime
and data security in the data center network.
By using the $RG$-factorizations of any absorbing Markov process,
we show that the two methods can be very effective in file lifetime analysis
of more general data center networks with file replication mechanism.
Finally, we use numerical examples to indicate impact of the threshold $d$
on the file average lifetime in this data center network.

The remainder of this paper is organized as follows. Section 2 describes
the data center network with file replication mechanism. Section 3 develops
two relevant Markov processes to give an approximate method
for assessing file average lifetime. Section 4 establishes
a QBD process to propse a two-dimensional method
for assessing file average lifetime, and uses a numerical example
to indicate impact of the threshold $d$ on the file average lifetime
in this data center network.

\section{Model Description}

In this section, we describe a large-scale data center network with file replication mechanism, in which each data center
may fail and new data centers can join this network, and a file $d$-replication policy is
proposed to increase the file lifetime and data security in the data center network.

For a large-scale data center network, we describe its physical structure, main random factors and
system parameters as follows:

\textbf{(1) The physical structure:} There are many data centers distributed in parallel a physical network with
different departments and/or geographical environment. For simplicity of analysis, we assume
that all the data centers are identical and are operated independently.

\textbf{(2) The lifetime:} Each data center in this network may be failure.
We assume that the lifetime $X$ of each data center follows an exponential distribution
with failure rate $\lambda >0$, that is, $P\left\{  X\leq t\right\}  =1-e^{-\lambda t}$.
Obviously, $E\left[X\right]  =1/\lambda$. If there are $k$ data centers in this network, then the failure
rate of the data center network is $k\lambda$ due to the exponential lifetime of each data center.

\textbf{(3) A joining process of new data centers:} Since the data center networks not only are fast
developing in the last over ten years but also each data center may be failure, new data centers
need to continually join to the network such that the data center network
can maintain a development of sustainability through many incessant equipment replacements.
We assume that the inputs of new data centers to the data centers network is a poisson
process with arrival rate $\beta >0$.

\textbf{(4) A file $d$-replication policy:} We assume that each file is stored
in at most $d\geq 1$ data centers in this network. Once the copy number of the
file is less than $d$ and there also exists an available data center without storing the file,
then the file will fast replicated to the data center.
We assume that the copy time $Y$ of the file replicated to the available data center
follows an exponential distribution with copy rate $\mu >0$, that is, $P\left\{  Y\leq t\right\} 
=1-e^{-\mu t}$.
Obviously, $E\left[  Y\right]  =1/\mu$. If there are $k$ identical copy files be being duplicated to $k$
different data centers, then the copy time distribution of the $k$ identical copy files is exponential
with copy rate $k\mu$.

\textbf{(5) The file lost process:} Once a data center fails, then all its files and
useful information in the data center will be lost immediately.

We assume that all the random variables involved in the data center
network are independent of each other.

\section{An Approximate Assessment Method of File Lifetime}

In this section, we first set up a birth-death process to study the steady-state
probability distribution of the number of available data centers in the data
center network. Then we establish another birth-death process to give an
approximate assessment method of file lifetime in the the data
center network with file $d$-replication policy.

\subsection{The number of available data centers}

In this data center network, each data center may fail, and its lifetime
of staying in the network follows an exponential distribution with failure rate $\lambda$.
On the other hand, new data centers continuously join to the data
center network, and their inputs are a Poisson process with arrival rate
$\beta$.

Let $N(t)$ be the number of available data centers normally operating in the
data center network at the time $t$. Then $\left\{  N(t):t\geq0\right\}  $ is
a birth-death process on state space $\Omega=\left\{  0,1,2,...\right\}  $ whose
state transition relation is shown in Figure 1 .

\begin{figure}[ptb]
\setlength{\abovecaptionskip}{0cm}  \setlength{\belowcaptionskip}{-0cm}
\centering            \includegraphics[width=8cm]{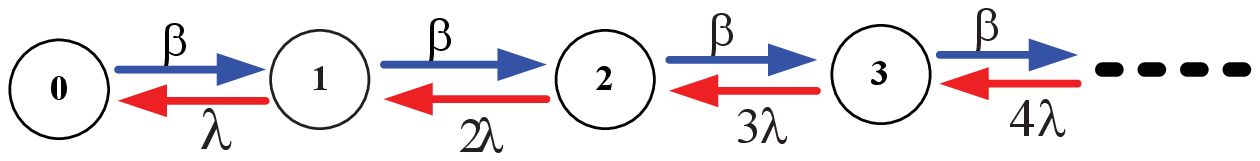}  \newline
\caption{State transition relation of a birth-death process}%
\label{figure:figure-1}%
\end{figure}

From Figure 1, the infinitesimal generator of the birth and death process $\left\{  N(t):t\geq0\right\}$
is given by
\[
Q=\left(
\begin{array}
[c]{ccccc}%
-\beta & \beta &  &  & \\
\lambda & -(\lambda+\beta) & \beta &  & \\
& 2\lambda & -(2\lambda+\beta) & \beta & \\
&  & \ddots & \ddots & \ddots
\end{array}
\right).
\]
Since the two numbers $\lambda$ and $\beta$ are fixed, there must exist a positive integer $n_0 > \lfloor\beta/\lambda\rfloor +1$
such that when $n > n_0$, we have $\lambda n>\beta$, where $\lfloor x \rfloor$ is the maximal integer
part of the real number $x$. Thus, by using the mean-draft condition, it is easy to see that the
birth-death process $\left\{  N(t):t\geq0\right\}  $ is irreducible, aperiodic and positive recurrent.

Let%
\[
\mathbf{N}=\underset{t\rightarrow+\infty}{\lim}N\left(  t\right)  ,
\]%
\[
\theta_{k}=P\left\{  \mathbf{N}=k\right\}  ,k=0,1,2,\ldots,
\]%
\[
\theta=\left(  \theta_{0},\theta_{1},\theta_{2},\theta_{3},\ldots\right)  .
\]
Then it is clear that $\theta Q=0,\theta e=1$, where $e$ is a column vector with each
element one.

\begin{The}
\label{The:Poisson}In this data centers network, the steady state number $\mathbf{N}$ of available data centers
operating normally follows a Poisson distribution with parameter $\beta/\lambda$, that is%
\[
\theta_{k}=\exp\left\{  -\frac{\beta}{\lambda}\right\}  \frac{1}{k!}\left(
\frac{\beta}{\lambda}\right)  ^{k}, \ \ \ \ k=0,1,2,\ldots.
\]
\end{The}

\textbf{Proof:} By solving the linear equations $\theta Q=0,\theta e=1$, we get%
\[
\theta_{k}=\exp\left\{  -\frac{\beta}{\lambda}\right\}  \frac{1}{k!}\left(
\frac{\beta}{\lambda}\right)  ^{k}, \ \ \ k=0,1,2,\ldots,
\]
Thus the steady state number $\mathbf{N}$ of available data centers
operating normally follows a Poisson distribution with parameter $\beta/\lambda$.
This completes the proof. $\Box$

It is seen from Theorem 1 that the Poisson random variable $\mathbf{N}$ provides
useful information to understand the file $d$-replication policy. For example,
the probability that no file can be successfully duplicated to
a data center is given by $P\left\{  \mathbf{N}=0\right\}
=\exp\left\{ -\beta / \lambda\right\}$%

\subsection{An approximate assessment for file lifetime}

In this subsection, we first correct the replication rate of the files to the data
center network by means of the steady state (Poisson) probability $\theta_{k}$ for
$k\geq 0$. Then we establish a new finite-state birth-death process to
provide an approximate assessment for the file lifetime.

By using Theorem 1 and
\[
\theta_{k}=\exp\left\{  -\frac{\beta}{\lambda}\right\}  \frac{1}{k!}\left(
\frac{\beta}{\lambda}\right)  ^{k}, \ \ \ k=0,1,2,\ldots,
\]
thus we can correct the replication rate of the files to the data centers as follows:

\textbf{(a) }If there is only a file in the data center network, then the file
has the replication rate to another data center, given by%
\[
\mu_{1}=\mu\sum\limits_{j=2}^{\infty}\theta_{j}=\mu\sum\limits_{j=2}^{\infty
}\exp\left\{  -\frac{\beta}{\lambda}\right\}  \frac{1}{j!}\left(  \frac{\beta
}{\lambda}\right)  ^{j}.
\]
That is, the copying time of this file follows an exponential distribution with copy rate $\mu_{1}$.

\textbf{(b) }If there are $k$ identical copy files in the data center network for $2\leq k\leq d-1$,
then the $k$ file has the replication rate to another data center, given by%
\[
\mu_{k}=k\mu\sum\limits_{j=k+1}^{\infty}\theta_{j}=k\mu\sum\limits_{j=k+1}%
^{\infty}\exp\left\{  -\frac{\beta}{\lambda}\right\}  \frac{1}{j!}\left(
\frac{\beta}{\lambda}\right)  ^{j}.
\]
That is, the copying time of the $k$ identical copy files follows
an exponential distribution with copy rate $\mu_{k}$.

In the data center network, we denote by $M(t)$ the number of identical copy
files of one file at the time $t$, then $\left\{ M(t): t\geq0\right\}$ is a birth-death
process on a finite state space $\mathbf{E=}\left\{  0,1,2,\ldots,d-1,d\right\}$
whose state transition relation is depcited in Figure 2.

\begin{figure}[ptb]
\setlength{\abovecaptionskip}{0cm}  \setlength{\belowcaptionskip}{-0cm}
\centering            \includegraphics[width=8cm]{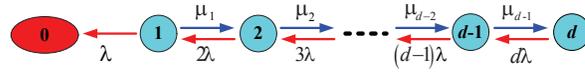}  \newline
\caption{State transition relation of a file replicated process}%
\label{figure:figure-2}%
\end{figure}

Let%
\[
\eta=\inf\left\{  t\geq0:M\left(  t\right)  =0\right\}  ,
\]
Then $\eta$ is the lifetime of a file which stays in the data center network. Of course, it is also
the first lost time of the file which will possibly disappear in the data center network.

We write
\[
S=\left(
\begin{array}
[c]{ccccc}%
-(\lambda+\mu_{1}) & \mu_{1} &  &  & \\
2\lambda & -(2\lambda+\mu_{2}) & \mu_{2} &  & \\
& \ddots & \ddots & \ddots & \\
&  & (d-1)\lambda & -\left(  (d-1)\lambda+\mu_{d}\right)  & \mu_{d}\\
&  &  & d\lambda & -d\lambda
\end{array}
\right)  ,S^{0}=\left(
\begin{array}
[c]{c}%
\lambda\\
0\\
\vdots\\
0\\
0
\end{array}
\right)  .
\]
Also, we take the initial probability $(\widetilde{\gamma}, \gamma_{0})$, where $\widetilde{\gamma}%
=(\gamma_{1},\gamma_{2},\ldots,\gamma_{d})$, $\gamma_{0}\in\left[  0,1\right]
$, and $\widetilde{\gamma}e=1-\gamma_{0}$.

\begin{The}
\label{The:PH distribution}In this data center network, the lifetime
$\eta$ of a file follows a PH distribution of size $d$ with an irreducibility
representation $(\widetilde{\gamma}, S)$. Also, the Markov process $\left(
S+S^{0}\widetilde{\gamma}\right) $ is irreducible. Further, the $k$th moment of
the lifetime $\eta$ of a file is given by%
\[
E\left[  \eta^{k}\right]  =\left(  -1\right)  ^{k}k!\widetilde{\gamma}%
S^{-k}e, \ \ \ k=1,2,3,\ldots.
\]
\end{The}

\textbf{Proof:} It is easy to check that the infinitesimal generator of the birth-death process
$\left\{  M(t): t\geq0\right\}  $ on state space $\mathbf{E=}\left\{
0,1,2,\ldots,d-1,d\right\}  $ is given by%
\[
\mathbf{Q=}\left(
\begin{array}
[c]{cc}%
S & S^{0}\\
0 & 0
\end{array}
\right)  .
\]
Obviously, the lifetime $\eta$ of a file follows a PH distribution of size $d$ with an irreducibility
representation $(\widetilde{\gamma}, S)$. Also, the Markov process $\left(
S+S^{0}\widetilde{\gamma}\right) $ is irreducible. In addition, some simple
computation can lead to the $k$th moment of the lifetime $\eta$. This completes the proof. $\Box$

Note that the matrix $S$ is the infinitesimal generator of a birth-death process, thus we can give
expression for the inverse of matrix $S$. To this end, we write%
\[
S^{-1}=\left(
\begin{array}
[c]{ccccc}%
S_{1,1} & S_{1,2} & \cdots & S_{1,d-1} & S_{1,d}\\
S_{2,1} & S_{2,2} & \cdots & S_{2,d-1} & S_{2,d}\\
\vdots & \vdots &  & \vdots & \vdots\\
S_{d-1,1} & S_{d-1,2} & \cdots & S_{d-1,d-1} & S_{d-1,d}\\
S_{d,1} & S_{d,2} & \cdots & S_{d,d-1} & S_{d,d}%
\end{array}
\right)  ,
\]
It is easy check from $SS^{-1}=I$ that the first column of $S^{-1}$\ is given by%
\[
s_{j,1}=-\frac{1}{\lambda}, \ \ \ \ 1\leq j\leq d,
\]
and for $2\leq k\leq d$, the $k$th column of \ $S^{-1}$ is given by%
\begin{align*}
s_{1,k}  &  =-\frac{\prod\limits_{j=1}^{k-1}\mu_{j}}{k!\lambda^{k}}%
,s_{2,k}=-\frac{\left(  \lambda+\mu_{1}\right)  \prod\limits_{j=2}^{k-1}%
\mu_{j}}{k!\lambda^{k}},s_{3,k}=-\frac{2!\lambda^{2}+\left(  \lambda+\mu
_{1}\right)  \prod\limits_{j=3}^{k-1}\mu_{j}}{k!\lambda^{k}},\\
\cdots,s_{k-1,k}  &  =-\frac{\left[  \left(  k-2\right)  !\lambda^{k-2}+\left(
k-3\right)  !\lambda^{k-3}\mu_{k-2}+\cdots+\lambda\prod\limits_{j=2}^{k-2}%
\mu_{j}+\prod\limits_{j=1}^{k-2}\mu_{j}\right]  \mu_{k-1}}{k!\lambda^{k}},
\end{align*}
and for $k\leq j\leq d$,%
\[
s_{j,k}=-\frac{\left(  k-1\right)  !\lambda^{k-1}+\left(  k-2\right)
!\lambda^{k-2}\mu_{k-1}+\cdots+\lambda\prod\limits_{j=2}^{k-1}\mu_{j}%
+\prod\limits_{j=1}^{k-1}\mu_{j}}{k!\lambda^{k}}.%
\]
Thus we obtain
\[
E\left[  \eta\right]  =-\sum_{j=1}^{d}\sum_{i=1}^{d}\gamma_{i}s_{i,j}.
\]

\section{A Two-Dimensional Assessment of File Lifetime}

In this section, we establish a two-dimensional Markov process by means of
the number of available data centers and the number of identical copy
files of one file. Based on this, we propose a two-dimensional assessment method
of file lifetime in the data center network.

In the data center network, as seen above, let $N(t)$ and $M(t)$ be the numbers of
available data centers and of identical copy files of one file at the time $t$, respectively. 
Obviously, $N(t)\in\left\{  0,1,2,\ldots\right\}$ and $M(t)\in\left\{  0,1,2,\ldots
,d\right\}$. It is seen from the exponential and Poisson assumptions that
 $\left\{  N(t),M(t): t\geq0\right\}$
is a two-dimensional Markov process, and further a QBD process, whose state
transition relation is depicted in Figure 3.

\begin{figure}[ptb]
\setlength{\abovecaptionskip}{0cm}  \setlength{\belowcaptionskip}{-0cm}
\centering            \includegraphics[width=8.5cm]{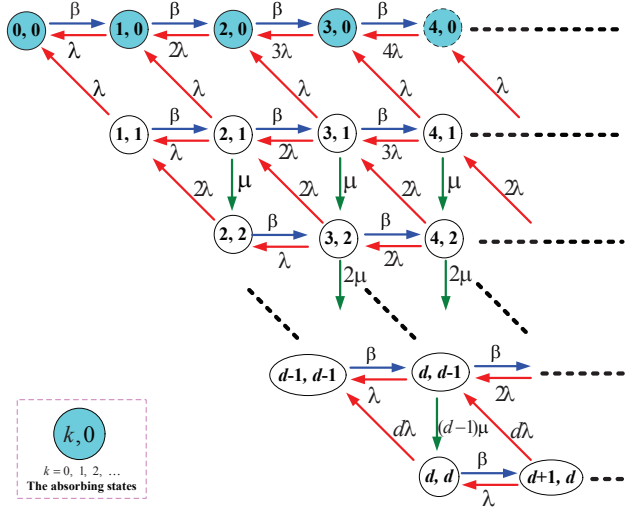}  \newline
\caption{State transition relation of a QBD process}%
\label{figure:figure-3}%
\end{figure}

It is seen from Figure 3 that the state space of the QBD process $\left\{
N(t),M(t): t\geq0\right\}$ is expressed as%
\[
\Theta=\Delta\cup\Theta_{1}\cup\Theta_{2}\cup\Theta_{3}\cup\cdots=\Delta
\cup\left(  \bigcup\limits_{k=1}^{\infty}\Theta_{k}\right),
\]
where $\Delta=\left\{  \left(  k,0\right)  :k=0,1,2\ldots
\right\}$ is a set of all the absorption states, which are written as
an absorbing state $\mathbf{\Delta}^{\ast}$. Observing the columns in 
Figure 3, we write

Level $k\in\left\{  1, 2, 3, \ldots, d-1\right\} : \Theta_{k}=\left\{  \left(
k,1\right)  ,\left(  k,2\right)  ,\ldots,\left(  k,k\right)  \right\}$;

Level $l\in\left\{  d, d+1, d+2, \ldots\right\} : \Theta_{l}=\left\{  \left(
l,1\right)  ,\left(  l,2\right)  ,\ldots,\left(  l,d\right)  \right\}  .$

From these levels, the infinitesimal generator of the
QBD process $\left\{  N(t),M(t): t\geq0\right\}  $ on sub-state space
$\bigcup_{k=1}^{\infty}\Theta_{k}$ is given by%
\[
T=\left(
\begin{array}
[c]{ccccc}%
A_{1,1} & A_{1,2} &  &  & \\
A_{2,1} & A_{2,2} & A_{2,3} &  & \\
& A_{3,2} & A_{3,3} & A_{3,4} & \\
&  & \ddots & \ddots & \ddots
\end{array}
\right)
\]
where $\zeta_{j}\left(  k\right)  =k\lambda+\beta+j\mu$, and
\[
A_{1,1}=-\left(  \lambda+\beta\right)  ,\text{ }A_{1,2}=\left(  \beta
,0\right);
\]
for $2\leq k\leq d$,
\[
A_{k,k-1}=\left(
\begin{array}
[c]{ccccc}%
\left(  k-1\right)  \lambda &  &  &  & \\
2\lambda & \left(  k-2\right)  \lambda &  &  & \\
& 3\lambda & \left(  k-3\right)  \lambda &  & \\
&  & \ddots & \ddots & \\
&  &  & \left(  k-1\right)  \lambda & \lambda\\
&  &  & 0 & k\lambda
\end{array}
\right),
\]%
\[
A_{k,k}=\left(
\begin{array}
[c]{ccccc}%
-\zeta_{1}\left(  k\right)  & \mu &  &  & \\
& -\zeta_{2}\left(  k\right)  & 2\mu &  & \\
&  & \ddots & \ddots & \\
&  &  & -\zeta_{k-1}\left(  k\right)  & \left(  k-1\right)  \mu\\
&  &  &  & -\zeta_{0}\left(  k\right)
\end{array}
\right), \ \ \
A_{k,k+1}=\left(
\begin{array}
[c]{cccccc}%
\beta &  &  &  &  & \\
& \beta &  &  &  & \\
&  & \beta &  &  & \\
&  &  & \beta &  & \\
&  &  &  & \beta & 0
\end{array}
\right);
\]%
and for $l\geq d+1$,%
\[
A_{l,l-1}=\left(
\begin{array}
[c]{ccccc}%
\left(  l-1\right)  \lambda &  &  &  & \\
2\lambda & \left(  l-2\right)  \lambda &  &  & \\
& 3\lambda & \left(  l-3\right)  \lambda &  & \\
&  & \ddots & \ddots & \\
&  &  & \left(  d-1\right)  \lambda & \left(  l-d+1\right)  \lambda\\
&  &  & d\lambda & \left(  l-d\right)  \lambda
\end{array}
\right)  ,
\]%
\[
A_{l,l}=\left(
\begin{array}
[c]{ccccc}%
-\zeta_{1}\left(  l\right)  & \mu &  &  & \\
& -\zeta_{2}\left(  l\right)  & 2\mu &  & \\
&  & \ddots & \ddots & \\
&  &  & -\zeta_{d-1}\left(  l\right)  & \left(  d-1\right)  \mu\\
&  &  &  & -\zeta_{0}\left(  l\right)
\end{array}
\right), \ \ \
A_{l,l+1}=\left(
\begin{array}
[c]{cccccc}%
\beta &  &  &  &  & \\
& \beta &  &  &  & \\
&  & \beta &  &  & \\
&  &  & \beta &  & \\
&  &  &  & \beta & 0
\end{array}
\right).
\]%
Further, the infinitesimal generator of the QBD process $\left\{  N(t),M(t):t\geq0\right\}
$ on a modified state space $\mathbf{\Delta}^{\ast}\cup\left(  \bigcup
_{k=1}^{\infty}\Theta_{k}\right)$ is given by%
\[
\mathbf{Q}=\left(
\begin{array}
[c]{cc}%
T^{0} & T\\
0 & 0
\end{array}
\right),
\]
where%
\[
T^{0}=-Te=\left(  \lambda;\lambda,0;\lambda,0,0;\lambda,0,0,0;\lambda
,0,0,0,0;\ldots\right)^T,
\]
and $a^T$ represents the transpose of the row vector $a$ .

Let%
\[
\chi=\inf\left\{  t\geq0:M\left(  t\right)  =0,N\left(  t\right)  \in\left\{
0,1,2,\ldots\right\}  \right\}  ,
\]
Then the random variable $\chi$ is the first passage time that the
QBD process $\left\{N(t),M(t): t\geq0\right\}$ reaches the absorption state $\mathbf{\Delta}^{\ast}$
for the first time. That is, the random variable $\chi$ is the lifetime of a file of staying in data center network.

To use the PH distribution, we take an initial probability vector
$\alpha=\left(  \alpha_{\Delta^{\ast}},\alpha_{1},\alpha_{2},\alpha_{3}%
,\ldots\right)  $, and $\alpha_{\Delta^{\ast}}\in\left[
0,1\right]$. For $1\leq k\leq d$,%
\[
\alpha_{k}=\left(  \alpha_{k,1},\alpha_{k,2},\cdots,\alpha_{k,k-1}%
,\alpha_{k,k}\right),
\]
and for $l\geq d+1$,%
\[
\alpha_{l}=\left(  \alpha_{l,1},\alpha_{l,2},\cdots,\alpha_{l,d-1}%
,\alpha_{l,d}\right)  .
\]

\begin{The}
\label{The:PH}In this data center network, the first passage time $\chi$\ is
an infinite-dimensional PH distribution with an irreducible representation
$\left(  \widetilde{\alpha},T\right)  $, where $\widetilde
{\alpha}=\left(  \alpha_{1},\alpha_{2},\alpha_{3},\cdots\right)$,
$\widetilde{\alpha}e=1-\alpha_{\mathbf{\Delta}^{\ast}}$. Also, the Markov
Process $T+T^{0}\widetilde{\alpha}$\ is irreducible. Further, the $k$th
moment of the first passage time $\chi$ is given by%
\[
E\left[  \chi^{k}\right]  =\left(  -1\right)  ^{k}k!\widetilde{\alpha}%
T^{-k}e, \ \ \ k=1,2,3,\ldots.
\]
\end{The}

\textbf{Proof:} Corresponding to the modified state space
$\mathbf{\Delta}^{\ast}\cup\left(\bigcup_{k=1}^{\infty}\Theta_{k}\right)$,
the infinitesimal generator of the QBD process
$\left\{  N(t),M(t): t\geq0\right\}$ is given by $\mathbf{Q}$.
Thus it is clear that the first passage time $\chi$ is an infinite-dimensional
PH distribution with an irreducible representation
$\left(\widetilde{\alpha},T\right)$.
Also, the Markov Process $T+T^{0}\widetilde{\alpha}$\ is irreducible, and the $k$th
moment of the first passage time $\chi$ is also obtained. This completes the proof.
$\Box$

It is necessary to show that the $RG$-factorizations by Li \cite{Li:2010}
can be applied to effectively deal with the infinite-dimensional
PH distribution. Now, we calculate the mean $E\left[\chi \right]$.
To this end, we first need to derive the inverse matrix of the
matrix $T$ by using the $RG$-factorizations.

we define the $U-$measure as
\[
\mathbf{U}_{0}=A_{1,1},
\]%
and for $k=1,2,3,\ldots$,
\[
\mathbf{U}_{k}=A_{k+1,k+1}+A_{k+1,k}(-U_{k-1})^{-1}A_{k,k+1}.
\]
We respectively define the $R-$ and $G-$measures as%
\[
\mathbf{R}_{k}=A_{k+1,k}(-\mathbf{U}_{k-1})^{-1},\text{ }k=1,2,3,\ldots;
\]
\[
\mathbf{G}_{l}=(-\mathbf{U}_{l})^{-1}A_{l+1,l+2},\text{ }l=0,1,2,\ldots.
\]

By using the Theorem 1 in Section 2.1 of Li and Cao \cite{Li:2004}, 
the $RG$-factorization of matrix $T$ is given by%
\[
T=\left(  I-\mathbf{R}_{L}\right)  \mathbf{U}_{D}\left(  I-\mathbf{G}%
_{U}\right),
\]
where%
\[
\mathbf{U}_{D}=\text{diag}(\mathbf{U}_{0},\mathbf{U}_{1},\mathbf{U}_{2}%
,\ldots),
\]%
\[
\mathbf{R}_{L}=\left(
\begin{array}
[c]{ccccc}%
0 &  &  &  & \\
\mathbf{R}_{1} & 0 &  &  & \\
& \mathbf{R}_{2} & 0 &  & \\
&  & \mathbf{R}_{3} & 0 & \\
&  &  & \ddots & \ddots
\end{array}
\right)  ,\mathbf{G}_{U}=\left(
\begin{array}
[c]{ccccc}%
0 & \mathbf{G}_{0} &  &  & \\
& 0 & \mathbf{G}_{1} &  & \\
&  & 0 & \mathbf{G}_{2} & \\
&  &  & 0 & \ddots\\
&  &  &  & \ddots
\end{array}
\right)  .
\]
Thus we have obtain%
\[
T^{-1}=\left(  I-\mathbf{G}_{L}\right)  ^{-1}\mathbf{U}_{D}^{-1}\left(
I-\mathbf{R}_{L}\right)  ^{-1}.
\]
Let%
\[
X_{k}^{(l)}=\mathbf{R}_{l}\mathbf{R}_{l-1}\cdots\mathbf{R}_{l-k+1}, \ \ \ \ 1\leq k\leq l,
\]%
\[
Y_{k}^{(l)}=\mathbf{G}_{l}\mathbf{G}_{l+1}\cdots\mathbf{R}_{l+k-1}, \ \ \ \ 0\leq l\leq k,
\]
Then%
\[
\mathbf{U}_{D}^{-1}=\text{diag}(\mathbf{U}_{0}^{-1},\mathbf{U}_{1}^{-1},\mathbf{U}%
_{2}^{-1},\ldots),
\]%
and
\[
\left(  I-\mathbf{G}_{L}\right)  ^{-1}=\left(
\begin{array}
[c]{ccccc}%
I & Y_{1}^{\left(  0\right)  } & Y_{2}^{\left(  0\right)  } & Y_{3}^{\left(
0\right)  } & \cdots\\
& I & Y_{1}^{(1)} & Y_{2}^{(1)} & \cdots\\
&  & I & Y_{1}^{(2)} & \cdots\\
&  &  & I & \cdots\\
&  &  &  & \ddots
\end{array}
\right)  ,\left(  I-\mathbf{R}_{L}\right)  ^{-1}=\left(
\begin{array}
[c]{ccccc}%
I &  &  &  & \\
X_{1}^{(1)} & I &  &  & \\
X_{2}^{(2)} & X_{1}^{(2)} & I &  & \\
X_{3}^{(3)} & X_{2}^{(3)} & X_{1}^{(3)} & I & \\
\vdots & \vdots & \vdots & \vdots & \ddots
\end{array}
\right)  .
\]
Let%
\[
T^{-1}=\left(  I-\mathbf{G}_{L}\right)^{-1} \mathbf{U}_{D}^{-1} \left(
I-\mathbf{R}_{L}\right)^{-1}
=\left(
\begin{array}
[c]{cccc}%
T_{0,0} & T_{0,1} & T_{0,2} & \cdots\\
T_{1,0} & T_{11} & T_{1,2} & \cdots\\
T_{2,0} & T_{2,1} & T_{2,2} & \cdots\\
\vdots & \vdots & \vdots & \ddots
\end{array}
\right).
\]
Then the mean of the first passage time $\chi$ is given by%
\begin{align*}
E\left[  \chi\right]=-\sum\limits_{j=0}^{\infty}\sum\limits_{i=0}^{\infty}\alpha_{i+1}T_{i,j}e.
\end{align*}

Finally, we use a simple example to illustrate how the file
lifetime $\chi$ depends on the maximum number of identical backups: $d\in\left(
2,59\right)$. Let $\lambda=1$ and $\beta=4$. As seen form figure 4, the mean $E\left[  \chi\right]$ increases,
as $d$ increases. In addition, when $d$ increases to a certain value,
the mean $E\left[\chi\right]$ will no longer change significantly. Such a
phenomenon will be very useful in design and optimization of the data center network with file replication mechanism.

\begin{figure}[ptb]
\setlength{\abovecaptionskip}{0cm}  \setlength{\belowcaptionskip}{-0cm}
\centering            \includegraphics[width=8cm]{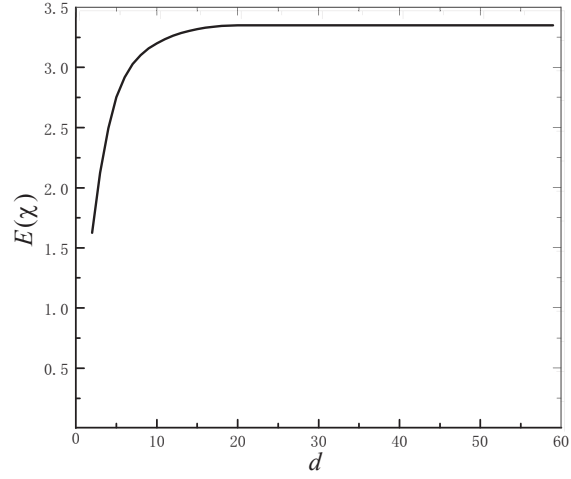}  \newline
\caption{The file lifetime depends on the key parameter $d$}%
\label{figure:figure-4}%
\end{figure}

%
%
%

\end{document}